\documentclass[twocolumn,epsfig,aps,prb]{revtex4}

\begin{document}

\title{ Entropy bounds in nonlinear quantum nanooptics }

\author{Igor I. Smolyaninov}
\affiliation{Department of Electrical and Computer Engineering, University of Maryland, College Park, MD 20742, USA}

\date{\today}

\begin{abstract}
Optical entropy bounds for metal nanoparticles immersed in nonlinear optical media and for nonlinear dielectric microdroplets on metal surfaces are calculated near the frequency of the surface plasmon resonance. Similar to the Bekenstein-Hawking result for the black hole entropy, the entropy bounds in nonlinear quantum nanooptics may be expressed as the ratios of the droplet perimeter (nanoparticle area) to the effective Planck length (effective Planck length squared). 
\end{abstract}


\maketitle

Linear and nonlinear nano-optics of surface plasmon polaritons (SPP) \cite{1,2} attracts great deal of current attention due to wide range of its applications in such diverse areas as optical nanolithography \cite{3}, sub-diffraction-limited microscopy \cite{4}, quantum computing \cite{5}, optoelectronics device integration on sub-micrometer scales \cite{6,7}, biosensing \cite{8}, etc. Many of these applications are possible due to the fact that the SPP wavelength in a given frequency range may be much shorter than the wavelength of free space photons. Such tight spatial localization of the SPP field gives rise to considerable field enhancement and hence, to many interesting nonlinear optical phenomena, which occur at very low light levels \cite{9,10}. As the number of SPP quanta involved in these effects is reduced, quantum mechanical effects play more and more important role in nonlinear nanooptics. Quantum sources of SPP or SPASERS \cite{11} have been theoretically proposed recently, while experimentalists move close to realization of such quantum nonlinear nano-optical devices \cite{12}. Thus, reaching good understanding of nonlinear quantum nano-optics is a very important current goal.
However, as we have pointed out recently \cite{13}, this is a highly nontrivial theoretical task. Its complexity may be illustrated by recalling the analogy between the nonlinear quantum nano-optics and the quantum gravity, which has been indicated in \cite{13}. 

It is well known (see for example ref.\cite{14} and the references therein) that the refractive index $n=(\epsilon \mu)^{1/2}$ plays the role of an effective metrics in optics. The Maxwell equations in a general curved space-time background $g_{ij}(x,t)$ are equivalent to the phenomenological Maxwell equations in the presence of a matter background with the following spatial distribution of the dielectric and magnetic permeability tensors $\epsilon _{ij}(x,t)$ and $\mu _{ij}(x,t)$ \cite{1}:

\begin{equation}
\label{eq1} 
\epsilon _{ij}(x,t)=\mu _{ij}(x,t)=(-g)^{1/2}\frac{ g^{ij}(x,t)}{g_{00}(x,t)}
\end{equation}
 
If we extend this analogy to nonlinear optics phenomena, we must conclude that the expression for the dielectric constant $\epsilon _d$ of a Kerr medium in nonlinear optics:

\begin{equation}
\label{eq2}
\epsilon _d= \epsilon ^{(1)} + 4\pi \chi ^{(3)}E^2 
\end{equation}

(where $E$ is the optical electric field, $\epsilon ^{(1)}$ and $\chi ^{(3)}$ are the linear dielectric constant and the third order nonlinear susceptibility of the medium, and for the sake of simplicity we omit tensor indices of $\epsilon ^{(1)}$ and $\chi ^{(3)}$) plays the same role as the role of the Einstein equation in general relativity:

\begin{equation}
\label{eq3} 
R_{ik}=\frac{8\pi k}{c^4}T_{ik} ,
\end{equation}

where $k$ is the gravitational constant, and we assume $T=0$. Both equations define the influence of the energy-momentum tensor on the effective metric. For evident reasons nonlinear optics cannot emulate gravity exactly. However, from comparison of equations (2) and (3) it is clear that $\chi ^{(3)}$ in nonlinear optics plays roughly the same role as the gravitational constant in eq.(3): positive $\chi ^{(3)}$ results in self-focusing or, in other words, attractive interaction of optical quanta. 

The spatio-temporal scale at which quantum metric fluctuations play an important role in gravitational effects is characterized by the Planck length $L_p$ (see ref. \cite{15} for a recent review on quantum gravity):

\begin{equation}
\label{eq4} 
L_p=(\frac{\hbar k}{c^3})^{1/2}\approx 1.6\times 10^{-33} cm 
\end{equation}

Due to extreme smallness of this parameter, effects of quantum gravity are out of experimental reach. On the other hand, the effective nonlinear-optical Planck scale originally introduced in \cite{13} does not appear that far out of reach. In order to avoid undesirable wave-vector dependence in the original expression from ref. \cite{13}, let us introduce the effective Planck scale as

\begin{equation}
\label{eq5} 
L^{eff}_p=(\hbar \chi ^{(3)}c)^{1/4}
\end{equation}
 
This expression is based on simple dimensional analysis. Assuming $\chi ^{(3)}\sim 10^{-10}$ esu, as in experiments with 4BCMU polidyacetylene films on gold in ref.\cite{9}, we obtain $L^{eff}_p\approx 2.3$ nm. While this value is still too small for usual nonlinear optical experiments, it plays an important role in quantum nonlinear optics of surface plasmon polaritons \cite{13}. In fact, the effective two-dimensional (2D) Planck scale as seen by the SPPs having frequencies near the frequency of the surface plasmon resonance \cite{2} is considerably larger. In order to understand this, we must recall the basic facts about the dispersion law of SPP (more detailed discussion can be found in \cite{2,13}).

Let us consider SPPs which propagate along an interface between a metal film and a third-order nonlinear dielectric, which is located on both sides of the film. In such a case the dispersion law can be written as \cite{13} 

\begin{equation}  
\label{eq6} 
k^2=\frac{\omega ^2}{c^2}\frac{\epsilon _d\epsilon _m(\omega )}{\epsilon _d+\epsilon _m(\omega)\pm 2\epsilon _de^{-kd}} ,
\end{equation}

where $\epsilon _m(\omega )$ is the frequency-dependent dielectric constant of the metal. If we assume that $\epsilon _m=1-\omega _p^2/\omega ^2$ is real (where $\omega _p$ is the plasma frequency of the lossless metal) and that $d$ is large, the dispersion law can be simplified as

\begin{equation}  
\label{eq7} 
k^2=\frac{\omega ^2}{c^2}\frac{\epsilon _d\epsilon _m(\omega )}{\epsilon _d+\epsilon _m(\omega)} 
\end{equation}

This dispersion law approaches asymptotically $\omega _{sp}=\omega _p/(1+\epsilon _d)^{1/2}$ at large wave vectors. The latter frequency corresponds to the surface plasmon resonance. The dispersion law (6) also diverges near $\omega _{sp}$ in the general case of a lossy metal film if the metal film thickness $d$ is small, so that the imaginary part of the term $\pm 2\epsilon _de^{-kd}$ compensates the imaginary part of $\epsilon _m$. In addition, pumping energy into the dielectric by optical or other means may create a dielectric with gain in a given frequency range \cite{12}, so that the imaginary part of $\epsilon _d$ may compensate losses in metal and $k$ again diverges near $\omega _{sp}$. Very-short wavelength (in the 10-50 nm range) plasmons with frequencies near $\omega _{sp}$ were observed in a number of near-field optical experiments \cite{16,17}. 

It is clear from eqs.(6,7) that around $\omega _{sp}$ (near $\epsilon _m(\omega)=-\epsilon _d$) both phase velocity $c_{pl}$ and group velocity of surface plasmons tend to zero, and the effective two-dimensional dielectric constant $\epsilon _{2D}^{(1)}=c^2/c^2_{pl}$ of the nonlinear dielectric diverges as seen by the plasmons:

\begin{equation}  
\label{eq8} 
\epsilon _{2D}^{(1)}=\frac{\epsilon _d\epsilon _m(\omega )}{\epsilon _d+\epsilon _m(\omega)} ,
\end{equation}

In a similar fashion, the effective (as seen by SPPs) two-dimensional nonlinear susceptibility $\chi _{2D}^{(3)}$ of the dielectric diverges near $\omega _{sp}$: small changes of the three-dimensional dielectric constant $\epsilon _d$ due to the $4\pi \chi ^{(3)}E^2$ term in equation (2) are perceived as very large changes of the effective two-dimensional dielectric constant (8) by surface plasmon-polaritons:

\begin{equation}  
\label{eq9} 
\chi _{2D}^{(3)}=\chi ^{(3)}\frac{\epsilon ^2_m}{(\epsilon _d+\epsilon _m(\omega))^2} ,
\end{equation}

As a result, the effective 2D Planck scale as seen by the surface plasmons 

\begin{equation}
\label{eq10} 
L^{eff}_{2D}=(\hbar \chi _{2D}^{(3)}c)^{1/4}
\end{equation}

is much larger then $L^{eff}_p$ from eq.(5). As evident from eq.(9), the numerical value of $L^{eff}_{2D}$ is defined by the imaginary part of $(\epsilon _d+\epsilon _m)$. 

Let us estimate the largest possible $L^{eff}_{2D}$. Below we will show that $L^{eff}_{2D}$ defines the large-wavevector cut-off of the SPP dispersion law in the lossless metal case. Thus, while setting up experimental conditions for excitation of short-wavelength SPP, we only need to compensate the imaginary part of $\epsilon _m$ to the extent that the smallest SPP wavelength achievable according to eqs.(6,7) is of the order of $L^{eff}_{2D}$. After straightforward calculations this condition gives the largest effective 2D Planck length achievable in the experiment as   

\begin{equation}
\label{eq11} 
L^{eff}_{2D}=(2^{1/2}L^{eff}_p\lambda_0)^{1/2}\approx 40 nm ,
\end{equation}

where $\lambda _0$ is the free space photon wavelength at the frequency of the surface plasmon resonance. This value is clearly of the same order of magnitude as the SPP wavelengths observed in the experiment, which means that the effective Planck scale is an important parameter in quantum nanooptics. The value of $L^{eff}_{2D}$ defines the thickness of the boundary layer of the nonlinear dielectric which exhibits the quantum highway mirage effect \cite{13}. In this surface boundary layer the average refractive index substantially exceeds the value of refractive index in the bulk (because of the Kerr shift caused by zero-point fluctuations of the SPP field). In addition, this boundary layer experiences very strong quantum fluctuations of the dielectric constant: $\Delta \epsilon _d\sim 1$ \cite{13}. As a result, surface plasmons, which propagate through this fluctuating dielectric layer experience strong dephasing (loss of phase coherence) effect. In fact, according to Ioffe-Regel criterium, SPP with a wavelength shorter than $L^{eff}_{2D}$ cannot propagate at all. Such SPPs experience Anderson localization. Thus, the value of $L^{eff}_{2D}$ defines the high momentum cut-off of the SPP dispersion law in a way that is similar to the momentum cut-off at $1/L_p$ in quantum gravity \cite{15}. 

Let us show that in many other respects the role of $L^{eff}_{2D}$ parameter in quantum nonlinear optics is very similar to the role of the Planck length in quantum gravity. First, let us show that it defines the quantum uncertainty of the dielectric constant in a way that is similar to the quantum gravitational uncertainty relation for the space-time metrics. Expressed in terms of connection $\Gamma$, the quantum gravitational uncertainty relation is $\Delta \Gamma l^3\geq L^2_{p}$, while in terms of the metric tensor it may be expressed as

\begin{equation}
\label{eq12} 
\Delta gl^2\geq L^2_p, 
\end{equation}

where $l$ is the linear size of a space-time region \cite{15}. This uncertainty in the gravitational field comes from the quantum fluctuations of space-time.

In the language of the effective metric (1), which is experienced by SPPs in their propagation along the metal-dielectric interface, this uncertainty relation is translated into 

\begin{equation}
\label{eq13} 
\frac{2\Delta\epsilon}{\epsilon^3}\geq\frac{L^{eff2}_{p}}{l^2}, 
\end{equation}

where $L_{p}^{eff}$ is obtained from eq.(5). It is easy to see that this expression coincides with eq.(8) from ref.\cite{13}, which was obtained via calculation of the zero-point fluctuations of the SPP field. Thus, the effective metric fluctuations experienced by SPPs are very strong: $\Delta g_{00}\sim g_{00}$, and the dephasing effect of the dielectric constant fluctuations described above is a direct analogue of the phase coherence loss in quantum gravity \cite{15}.

Below we will show that similar to the Bekenstein-Hawking result for the black hole entropy \cite{15}, the value of $L^{eff}_{2D}$ defines the optical entropy bounds for a metal nanoparticle immersed in a nonlinear optical medium, and for a nonlinear dielectric microdroplet on a metal surface. As has been shown in ref.\cite{18}, the effective metrics experienced by the SPPs near such objects may to some degree emulate the metrics of a real black hole. We will show that the entropy bounds in nonlinear quantum nanooptics may be expressed as the ratios of the droplet perimeter (nanoparticle area) to the effective Planck length (effective Planck length squared). These results reproduce the well known quantum gravity expressions for the entropy of real black holes. 

Let us briefly recall the surface plasmon black hole analogy, which is described in detail in \cite{18}. If a dielectric object (say, a droplet) is placed on the metal surface (see Fig.1), the SPP dispersion law will be a function of the local thickness of the droplet and the local value of its dielectric constant, which may or may not be constant throughout the droplet. We may assume that the spatial distribution of the dielectric constant $\epsilon _d$ inside the droplet is chosen by an experimentalist such that at each illumination frequency in the range between $\omega _p/(1+\epsilon _d)^{1/2}$ and $\omega _p/2^{1/2}$ (these are the frequencies of the SP resonance at the metal-dielectric and metal-vacuum interfaces, respectively) there will be a closed linear boundary inside the droplet for which the SP resonance conditions are satisfied, and both phase and group velocity of surface plasmons tend to zero at this boundary. Quantitatively the effective metric experienced by surface plasmons near the droplet boundary may be written as

\begin{equation}
\label{eq14}
ds^2=c^{\star 2}dt^2-dx^2-dy^2 ,
\end{equation}

where $c^\star (x,y)$ is the local SPP phase velocity \cite{18}. The behavior of $c^\star (x,y)$ is defined by the shape and thickness of the droplet near its edge, by the thickness of the metal film, and by the frequency of SPPs. In order to emulate the Rindler metric, the droplet geometry may be chosen so that $c^\star =\alpha xc$ in the vicinity of $x=0$. Naturally, only some features of real black holes are emulated in this model. The limitations of this model are described in detail in \cite{18}.  

This toy-model is also limited in terms of emulating only the motion of individual electromagnetic quanta in some static effective metric, which is unaffected by the motion and changing spatial distribution of other electromagnetic quanta in the system. This limitation may be lifted if we consider an experiment performed with a droplet made of nonlinear dielectric material. If the optical nonlinearity of the droplet may be described by equation (2) with $\chi ^{(3)}>0$, the self-focusing of surface plasmons may lead to an effective "gravitational collapse" of the SPP field near the critical surface inside the droplet. As described above, this type of nonlinearity causes an effective attractive interaction of SPPs with each other. Thus, we may imagine a situation in which a liquid droplet is illuminated with an intense plasmon beam at a frequency below $\omega _p/(1+\epsilon ^{(1)}_d)^{1/2}$, so that a low intensity SPP field would not experience a critical surface near the droplet edge. However, the increase in the droplet refractive index due to the high intensity SPP field will cause the appearance of the critical surface at which the surface plasmon resonance conditions are satisfied. As a result, the plasmon field will collapse towards this arising critical surface. In order to complete the picture, we should take into account the strong quantum metric fluctuations near the emerging critical surface. Similar to the case of metal-dielectric boundary described in \cite{13}, the average refractive index of this thin fluctuating layer (with the thickness of the order of $L^{eff}_{2D}$) will be larger than the refractive index of the dielectric far from the critical surface. Thus, it would act as a self-induced waveguide and support SPP whispering gallery modes, which were experimentally observed in \cite{18}.

Let us compute the optical part of the entropy $S$ of such a toy SPP black hole/dielectric droplet produced as a result of the described SPP self-focusing experiment. We will do calculations in a usual way as $S=ln\Gamma $, where $\Gamma $ is the statistical sum \cite{19}. The dominant contribution to $\Gamma $ is produced by SPPs near the frequency of surface plasmon resonance. For such SPPs the dispersion law is almost flat. While they differ in their wave vectors, all the SPPs have the same frequency. That is why a very large entropy state is created in experiments near $\omega _{sp}$. If we assume that we have $N$ plasmons in such system, which are somehow distributed over $M$ possible single-plasmon states in the boundary layer adjacent to the critical surface (see Fig.1), the statistical sum is given by 

\begin{equation}
\label{eq15}
\Gamma =\frac{(M+N-1)!}{N!(M-1)!}\approx(\frac{M+N}{2\pi NM})^{1/2}\frac{(M+N)^{M+N}}{N^NM^M}
\end{equation}
   
Since the experiment is conducted near the frequency of the SP resonance, the sum $(\epsilon _d+\epsilon _m)$ is small. According to eq.(2) the maximum number of plasmons in the system is limited by the value of

\begin{equation}
\label{eq16}
E^2_{max}=\frac{-(\epsilon _m+\epsilon _d)}{4\pi\chi^{(3)}} 
\end{equation}

(note that the value of $\epsilon _d$ near the critical boundary is shifted due to zero-point fluctuations of the SPP field). Thus, necessarily $M>>N$ near the critical surface, and the statistical sum from eq.(15) may be simplified as $\Gamma \approx (M/N)^N$. As a result, we obtain a simple expression for the entropy as $S=Nln(M/N)$. The entropy bound is defined by the largest number of plasmons $N_{max}$, which may be fitted into the boundary layer near the critical surface, and thus by $E^2_{max}$ from eq.(16). As a result, $N_{max}$ may be estimated as 

\begin{equation}
\label{eq17}
2\pi rL^{eff2}_{2D}\frac{\epsilon _dE^2_{max}}{8\pi }\approx N_{max}\hbar \omega _{sp} , 
\end{equation}

and (neglecting the factor $ln(M/N)$ and other numerical factors of the order of one) we obtain $S_{max}\sim r/L^{eff}_{2D}$, where $r$ is the radius of the critical surface inside the droplet. In a similar fashion, the optical entropy bound for a metal nanoparticle immersed inside a nonlinear medium may be found as $S_{max}\sim r^2/L^{eff2}_{2D}$, where $r$ is the radius of the nanoparticle. These estimates reproduce the well-known quantum gravitational limit on the entropy of a given spatial region established by Bekenstein and Hawking.
While important for the field of quantum nonlinear nanooptics, these bounds represent an interesting example of a physical situation in which the entropy bounds have very transparent physical origins, and are easy to calculate.

In conclusion, optical entropy bounds of metal nanoparticles immersed in nonlinear optical media and nonlinear dielectric microdroplets on metal surfaces have been calculated near the frequency of the surface plasmon resonance. Similar to the Bekenstein-Hawking result for the black hole entropy, the entropy bounds in nonlinear quantum nanooptics may be expressed as the ratios of the droplet perimeter (nanoparticle area) to the effective Planck length (effective Planck length squared).

This work has been supported in part by the NSF grant ECS-0304046.

\begin{figure}
\begin{center}
\end{center}
\caption{ Self-focusing of the surface plasmon field and quantum fluctuations of the dielectric constant produce a fluctuating critical surface near the droplet edge as seen by the surface plasmon polaritons. }
\end{figure}

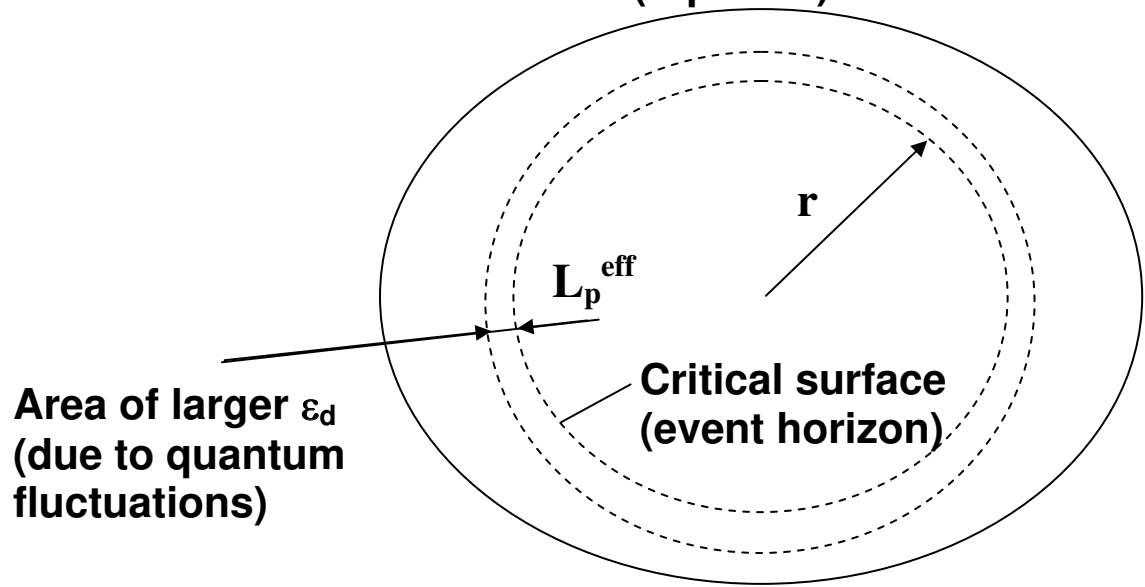
\end{document}